\documentclass[a4paper]{spie}  %>>> use this instead for A4 paper
\usepackage{amsmath,amsfonts,amssymb}
\usepackage[colorlinks=true, allcolors=blue]{hyperref}

\usepackage[english]{babel}
\usepackage[utf8]{inputenc}
\usepackage[]{graphicx}
\usepackage{microtype}
\usepackage{gensymb}
\usepackage{epsfig}
\usepackage{float}
\usepackage{gensymb}
\usepackage{todonotes}
\usepackage{boxedminipage2e}
\usepackage{siunitx}
\usepackage{booktabs}
\usepackage{wasysym}
\usepackage{subcaption}
\usepackage{todonotes}

\usepackage{natbib}
\bibpunct{(}{)}{;}{a}{}{,}

\graphicspath{{Pics/}}
\usepackage{txfonts}
\usepackage{enumerate}

\usepackage{pgfplots}
\pgfplotsset{compat=newest}
\usepgfplotslibrary{fillbetween}
\usetikzlibrary{backgrounds}

\title{The MICADO first light imager for the ELT: Eliminating vibrations and excessive earthquake-loads on specific subsystems} %instead of rotary platform

\author[a]{Nicklas, H.}
\author[a]{Schäfer, S.}
\author[a]{Anwand-Heerwart, H.}
\author[a]{Dette, J.-O.}
\author[a]{Witschel, J.}
\author[b]{Huber, H.} 
\affil[a]{Georg-August Universität Göttingen, Institut für Astrophysik und Geophysik, Friedrich-Hund-Platz\,1, 37077\,G\"ottingen,\,Germany}
\affil[b]{Max-Planck-Institute for Extraterrestrial Physics, Gießenbachstrasse\,1, 85748\:Garching,\:Germany}
\authorinfo{Further author information: (Send correspondence to Sebastian Sch\"afer)\\ Sebastian Sch\"afer: E-mail: schaefer@astro.physik.uni-goettingen.de, Telephone: +49 (0)551 39 25068}

\begin{document} 
%% Journal definitions borrowed from aa.cls

\def\aj{AJ}%
          % Astronomical Journal
\def\actaa{Acta Astron.}%
          % Acta Astronomica
\def\araa{ARA\&A}%
          % Annual Review of Astron and Astrophys
\def\apj{ApJ}%
          % Astrophysical Journal
\def\apjl{ApJ}%
          % Astrophysical Journal, Letters
\def\apjs{ApJS}%
          % Astrophysical Journal, Supplement
\def\ao{Appl.~Opt.}%
          % Applied Optics
\def\apss{Ap\&SS}%
          % Astrophysics and Space Science
\def\aap{A\&A}%
          % Astronomy and Astrophysics
\def\aapr{A\&A~Rev.}%
          % Astronomy and Astrophysics Reviews
\def\aaps{A\&AS}%
          % Astronomy and Astrophysics, Supplement
\def\azh{AZh}%
          % Astronomicheskii Zhurnal
\def\baas{BAAS}%
          % Bulletin of the AAS
\def\bac{Bull. astr. Inst. Czechosl.}%
          % Bulletin of the Astronomical Institutes of Czechoslovakia 
\def\caa{Chinese Astron. Astrophys.}%
          % Chinese Astronomy and Astrophysics
\def\cjaa{Chinese J. Astron. Astrophys.}%
          % Chinese Journal of Astronomy and Astrophysics
\def\icarus{Icarus}%
          % Icarus
\def\jcap{J. Cosmology Astropart. Phys.}%
          % Journal of Cosmology and Astroparticle Physics
\def\jrasc{JRASC}%
          % Journal of the RAS of Canada
\def\mnras{MNRAS}%
          % Monthly Notices of the RAS
\def\memras{MmRAS}%
          % Memoirs of the RAS
\def\na{New A}%
          % New Astronomy
\def\nar{New A Rev.}%
          % New Astronomy Review
\def\pasa{PASA}%
          % Publications of the Astron. Soc. of Australia
\def\pra{Phys.~Rev.~A}%
          % Physical Review A: General Physics
\def\prb{Phys.~Rev.~B}%
          % Physical Review B: Solid State
\def\prc{Phys.~Rev.~C}%
          % Physical Review C
\def\prd{Phys.~Rev.~D}%
          % Physical Review D
\def\pre{Phys.~Rev.~E}%
          % Physical Review E
\def\prl{Phys.~Rev.~Lett.}%
          % Physical Review Letters
\def\pasp{PASP}%
          % Publications of the ASP
\def\pasj{PASJ}%
          % Publications of the ASJ
\def\qjras{QJRAS}%
          % Quarterly Journal of the RAS
\def\rmxaa{Rev. Mexicana Astron. Astrofis.}%
          % Revista Mexicana de Astronomia y Astrofisica
\def\skytel{S\&T}%
          % Sky and Telescope
\def\solphys{Sol.~Phys.}%
          % Solar Physics
\def\sovast{Soviet~Ast.}%
          % Soviet Astronomy
\def\ssr{Space~Sci.~Rev.}%
          % Space Science Reviews
\def\zap{ZAp}%
          % Zeitschrift fuer Astrophysik
\def\nat{Nature}%
          % Nature
\def\iaucirc{IAU~Circ.}%
          % IAU Cirulars
\def\aplett{Astrophys.~Lett.}%
          % Astrophysics Letters
\def\apspr{Astrophys.~Space~Phys.~Res.}%
          % Astrophysics Space Physics Research
\def\bain{Bull.~Astron.~Inst.~Netherlands}%
          % Bulletin Astronomical Institute of the Netherlands
\def\fcp{Fund.~Cosmic~Phys.}%
          % Fundamental Cosmic Physics
\def\gca{Geochim.~Cosmochim.~Acta}%
          % Geochimica Cosmochimica Acta
\def\grl{Geophys.~Res.~Lett.}%
          % Geophysics Research Letters
\def\jcp{J.~Chem.~Phys.}%
          % Journal of Chemical Physics
\def\jgr{J.~Geophys.~Res.}%
          % Journal of Geophysics Research
\def\jqsrt{J.~Quant.~Spec.~Radiat.~Transf.}%
          % Journal of Quantitiative Spectroscopy and Radiative Trasfer
\def\memsai{Mem.~Soc.~Astron.~Italiana}%
          % Mem. Societa Astronomica Italiana
\def\nphysa{Nucl.~Phys.~A}%
          % Nuclear Physics A
\def\physrep{Phys.~Rep.}%
          % Physics Reports
\def\physscr{Phys.~Scr}%
          % Physica Scripta
\def\planss{Planet.~Space~Sci.}%
          % Planetary Space Science
\def\procspie{Proc.~SPIE}%
          % Proceedings of the SPIE
\let\astap=\aap
\let\apjlett=\apjl
\let\apjsupp=\apjs
\let\applopt=\ao

\maketitle

\begin{abstract}
Ensuring earthquake resilience is paramount for the upcoming Extremely Large Telescope (ELT). Instruments positioned on the ELT’s Nasmyth platform are anticipated to endure substantial seismic accelerations, peaking at 3.6\,g in specific scenarios. These instruments have to be designed not only to survive such events but also to only require minimal repairs for sustained optimal functionality. This paper presents our approach to mitigate the dynamic forces impacting the rotary platform. This platform hosts various critical components, including cryogenic control units and mosaic detector control systems, essential for the operation of MICADO in its mechanical de-rotating motion.

Furthermore, this paper tackles the challenge of minimizing instrument-induced vibrations affecting the telescope's structure. Given that the ELT’s optical system, composed of 798 segments, operates very similar to an interferometer, maintaining an extremely stable wavefront is crucial  - a key insight learned from the Very Large Telescope Interferometer (VLTI). Our focus is on identifying potential sources of vibration within specific frequency ranges where the telescope demonstrates increased sensitivity. We will explore strategies for mitigating these vibrations to maintain a wavefront error below 50\,nm.
\end{abstract}


% \section{ELT architecture}
% \input{01_ELT}
\vspace{2cm}

MICADO is the first light instrument for the upcoming ELT. In this paper we present the status of three main subsystems: Sec.\,\ref{sec:st0} shows the status of the main support structure, Sec.\,\ref{sec:rpa} illustrates the rotating platform and Sec.\,\ref{sec:ast} deals with the access structure. Furthermore, Sec.\,\ref{sec:vib} describes the impact of different devices on the vibrations transferred to the telescope and Sec.\,\ref{sec:eq} analyses how the potential for earthquakes had a strong impact on the design of the instrument. 

\section{Main Support Structure}
\label{sec:st0}
\begin{figure}[h]
    \centering
    \includegraphics[width=0.7\textwidth]{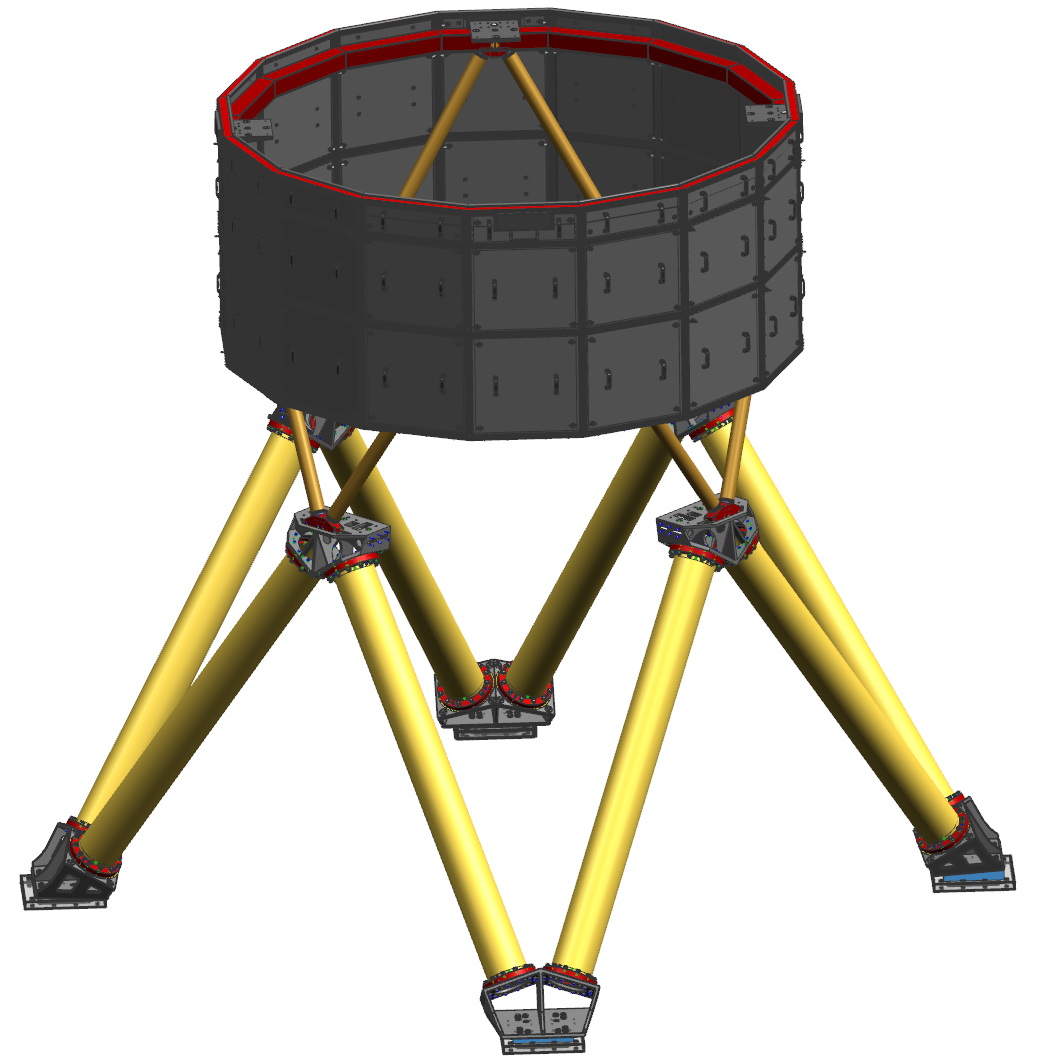}
    \caption{CAD drawing of MICADO's Main Support Structure.}
    \label{fig:ST0}
\end{figure}

MICADO's Main Support Structure (Fig.\,\ref{fig:ST0}) serves as a framework that carries most of the inner optical subsystems such as the cryostat and its de-rotator, SCAO, LOR and the relay optics (later to be replaced by MORFEO's M12) and keeps them in place, even in the event of an earthquake. For this purpose, it is required to provide the necessary mechanical interfaces with high precision and stability.

After successfully passing the final design review nearly a year ago, the MICADO support structure required only minor improvements in the spherical joints of the serrurier nodes. To facilitate easier alignment of the serrurier struts, the radius of the spherical faces was reduced, impacting the overall mass budget due to dimensional changes. Additionally, the spherical faces were moved radially outward, allowing the calotte fixation screws to run directly through the joint faces.

\subsection{Alignment Procedure and Strut Handling}
During testing and subsequent verification of the MICADO Support Structure, numerous iterations of serrurier strut alignment are anticipated initially. This is due to the large dimensions of the serrurier, where small positional and dimensional deviations caused by manufacturing tolerances significantly affect the positioning of the central and upper serrurier nodes. These re-alignment procedures necessitate shimming above and below the lower serrurier nodes. The resulting tensions are mitigated by adjusting strut lengths during shimming and allowing lateral movement in the strut-node joint by the two spherical parts.

\begin{figure}[ht]
    \centering
    \includegraphics[width = 0.8 \textwidth]{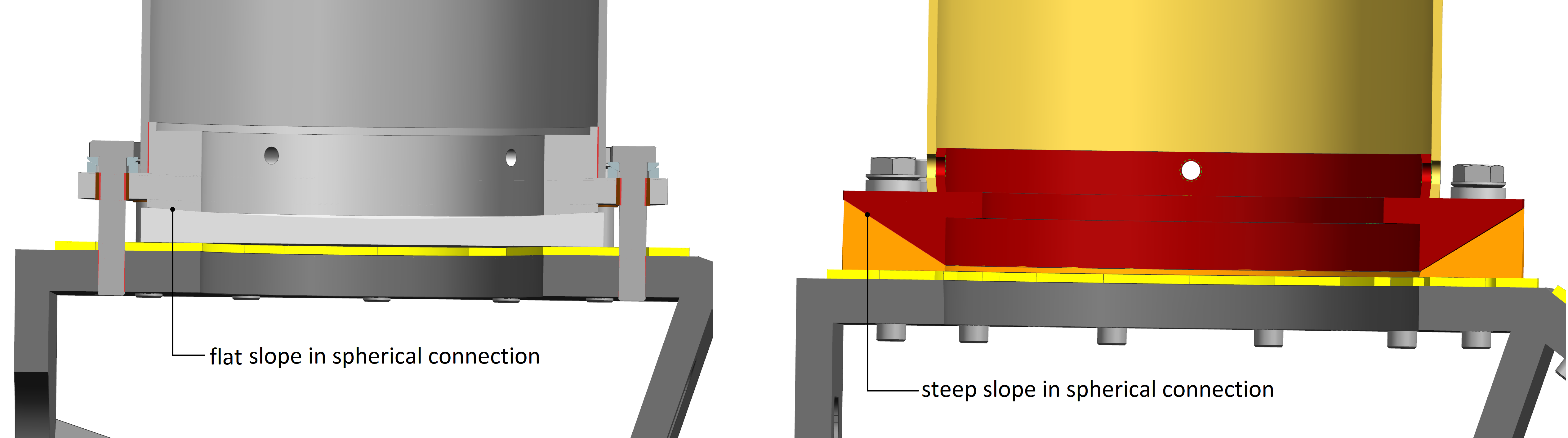}
    \caption{Spherical connections of support structure's serrurier nodes and struts: previous design (left) and final connection design (right).}
    \label{fig:ST1ab}
\end{figure}

The reduction in the spherical radii was based on the assumption that the male spherical part would settle more easily with a steeper 'hill,' enhancing the force pushing it into the female calotte disc center. Figure\,\ref{fig:ST1ab} visualizes the relevant design changes and impact on the spherical connection geometry.

\subsection{Side Effects and Mitigation}
The reduced sphere radius increased the weight of the spherical parts, resulting in a noticeable weight increase of approximately 300 kg, which is nearly 10 percent of the subsystem’s total mass budget (excluding contingency). This effect was amplified by the outward shift of the spherical faces, increasing the calotte ring diameter and thus the mass quadratically.

Weight optimizations halved this mass increase, primarily by adding a large opening in the center of the female spherical part, which was previously a closed surface. This modification allowed for a reduction in the total flange heights, transforming the calotte center from a physical component into a virtual point below the contact surface to the Serrurier node.

The newly introduced central opening enables more effective initial positioning, as the centering tool can now align both calotte discs with the Serrurier bottom node simultaneously, which was previously impossible. Additional push-off screws facilitate easier dismounting of the centering tool. The calotte flange, welded to the Serrurier strut, is now equipped with a step where the push-off screws engage and the next tool can be directly attached.

Due to the absence of material in the calotte disc center, the hydraulic jack cannot engage with the disc when lifting the strut assembly for shimming. This issue was resolved by introducing a removable cap against which the jack can press, thereby lifting the strut.

\section{Rotary Platform}
\label{sec:rpa}
\begin{figure}[h]
    \centering
    \includegraphics[width=1\textwidth]{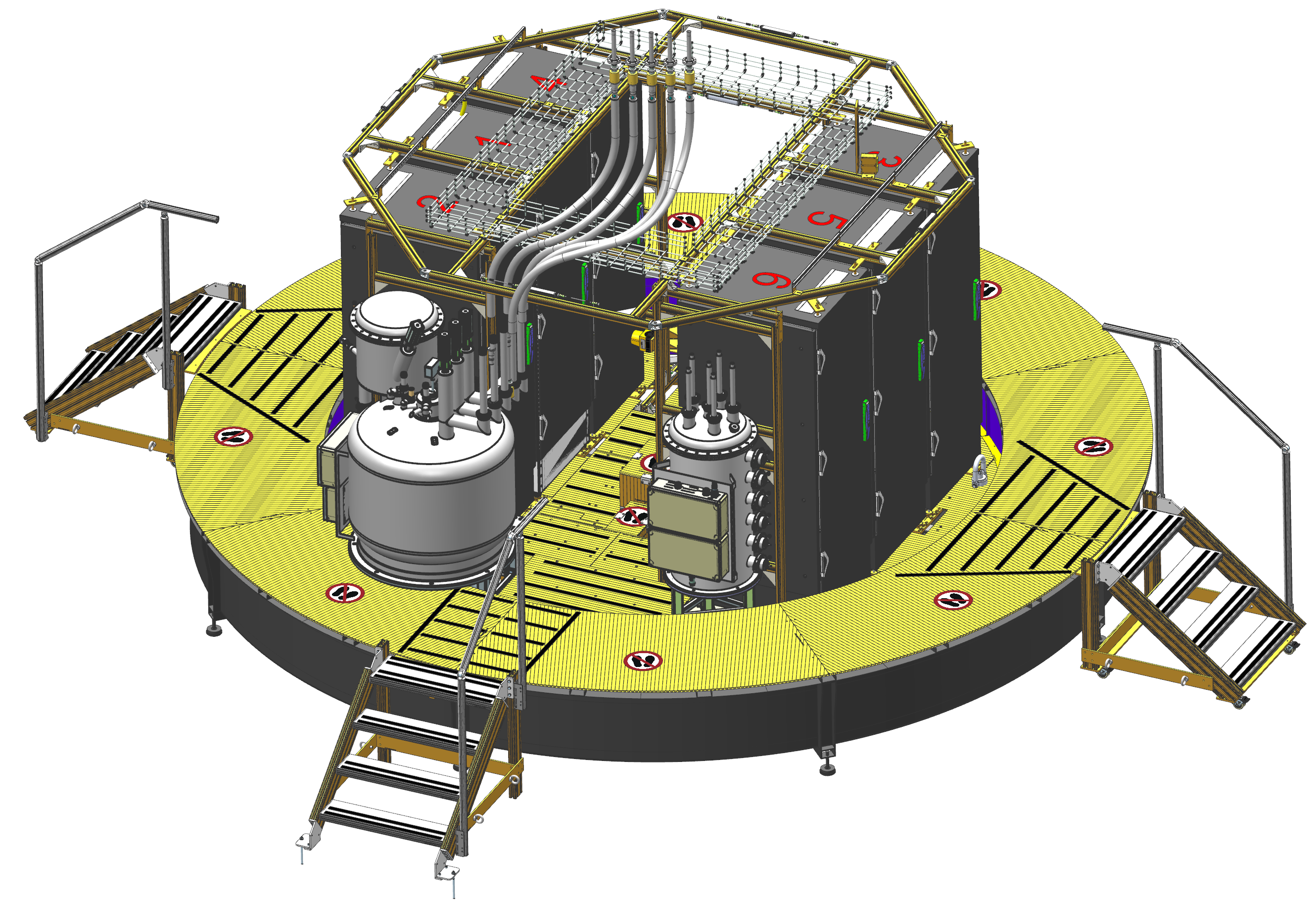}
    \caption{Rotary Platform isometric view. The central staircase is fixed while the outer two can be attached temporarily. }
    \label{fig:RPA1}
\end{figure}

The key components (detector, spectrometer, LOR, SCAO, etc.) of MICADO are positioned in or on top of the rotating de-rotator. It is essential that these critical components and their corresponding evaluation units are interconnected using short cable lengths. For this purpose, the co-rotating platform (RPA) follows the rotation of the instrument and simultaneously provides space for 6 cabinets, an LN2 buffer tank, and two backing pumps, as well as the water and gas supply (Fig.\,\ref{fig:RPA1}). Equipped with its own motor drive, it follows the movement independently. To monitor the movement, a safety switch, which controls the tracking of the Co-rotator in relation to the de-rotator, and an encoder for position control and a limit switch are installed. The platform is designed to be accessible for maintenance purposes.

\subsection{Cable Wrap}
The cable wrap system establishes the connection for power, cooling water, and data cables between the RPA and the fixed external components, such as the control cabinets on the Nasmyth platform. For this purpose, the cable wrap is positioned around the RPA and allows a rotation angle of +/-\,270\textdegree. The cables, with a length of approximately 12\,m, are guided in chains that the rotating platform drags along. Therefore, the cable wrap system does not need its own drive. The towing connection is "loose" so that no force is applied to the rotating platform. For control purposes, a safety sensor is integrated into the towing connection.

\subsection{Vibration Avoidance Concept}
The original solution for the RPA consisted of a central bearing on which an IPE frame was mounted. The IPE frame serves to hold the required cabinets. This IPE frame was supported at the outer ends by rollers to protect the central bearing from overloading in the event of an earthquake. During the assessment, however, it emerged that these rollers could also lift off in the event of an earthquake and then hit the track in an undefined manner. There were also concerns that the rollers could cause vibrations.

To resolve these concerns, we initially removed the rollers and the track from our concept. In order to protect the central bearing from overload in the event of an earthquake, it is necessary to minimize the deflection/stroke of the IPE frame at the outer ends of the IPE beams. As a solution, we have added an Earthquake Retaining System (ERS). This consists of a static ring that is mounted underneath the IPE frame on the Nasmyth platform. The outer ends of the frame are fitted with C-shaped components that enclose this ring and strike against the ring in the event of an earthquake, thus minimizing the deflection of the frame. As the ends of the IPE frame hang freely under normal conditions, this results in a deteriorated natural frequency. To improve the natural frequency, spring-mounted rollers running above and below the ERS ring were attached to the C-shaped components (see Fig.\,\ref{fig:RPA1}).

\begin{figure}[!h]
    \centering
    \includegraphics[width=1\textwidth]{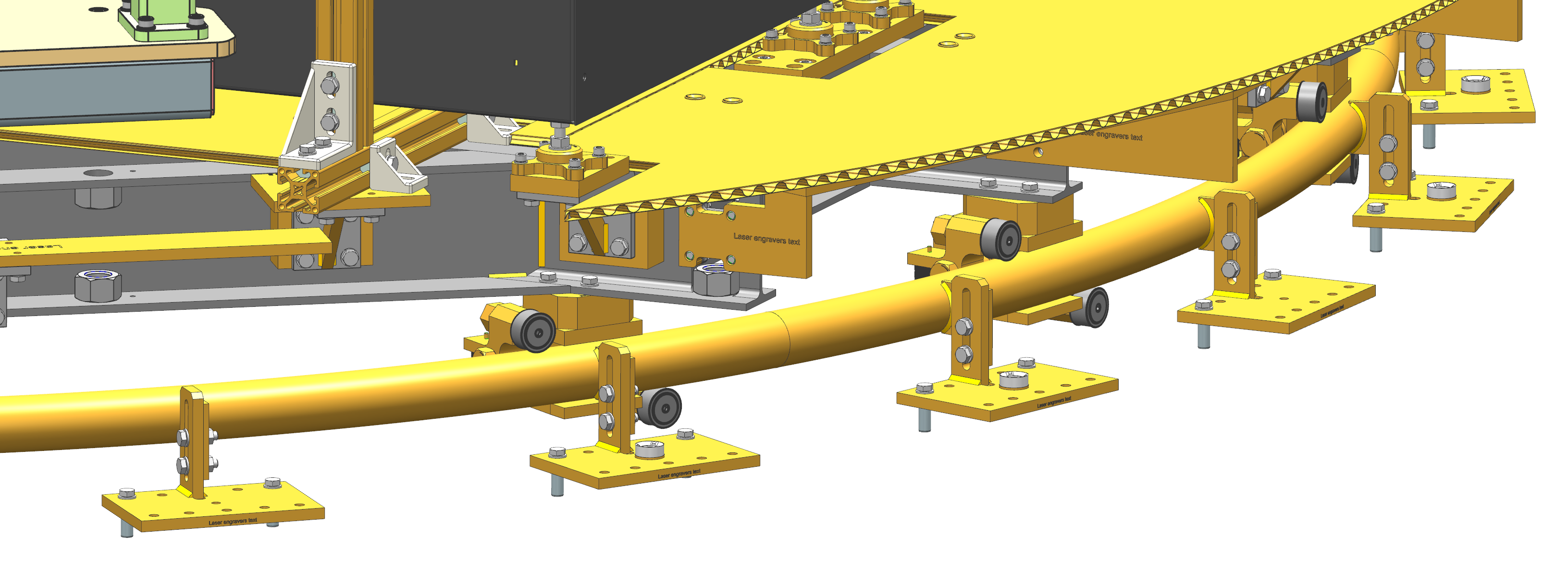}
    \caption{Rotary platform's retaining system unloading the central bearing from earthquake loads}
    \label{fig:RPA}
\end{figure}

\clearpage
\section{Access Structure Tool for maintenance}
\label{sec:ast}
MICADO's numerous maintenance cases demand extensive access to its subsystems, as well as to the relay optics and later to MORFEO’s M12 mirror atop the support structure. Access is required at three different height levels and ideally should be granted over a full 360\textdegree \,range. Limited available space on the Nasmyth platforms and highly demanding earthquake load cases drove the need for a dedicated custom-made scaffolding instead of off-the-shelf solutions such as scissor lifts or cherry pickers. Due to the geometric boundaries of the MICADO design volume and occasional interference with MORFEO maintenance activities, the access structure must possess a certain degree of flexibility to adapt to ever-changing constraints. This chapter provides an overview of the different influencing factors that drive and restrict the design of the access structure (AST). After demonstrating the resulting general layout partitioning, the various connection features are explained, allowing for modular configuration switches.

\subsection{Relay optics constraint}
As the only first-light instrument, MICADO will initially operate alone on ELT’s Nasmyth-A platform. As MORFEO will not be online initially, it will be temporarily substituted by the relay optics, which will be mounted atop MICADO. The installation’s overhanging optical bench prevents personnel from accessing parts of the access platform via regular walkways, necessitating an additional extension to grant radial access to the relay optic’s inner components. Maintenance staff working atop the bench are secured by hanging into an overarching safety gantry, which can only be mounted temporarily due to its additional weight. Access to the optical bench is provided by an additional stepladder mounted to the 2nd-floor maintenance platform (see Fig.\,\ref{fig:AST1} left).

\subsection{MORFEO servicing and constraints}
\begin{figure}[h]
\centering
\includegraphics[width=0.5\textwidth]{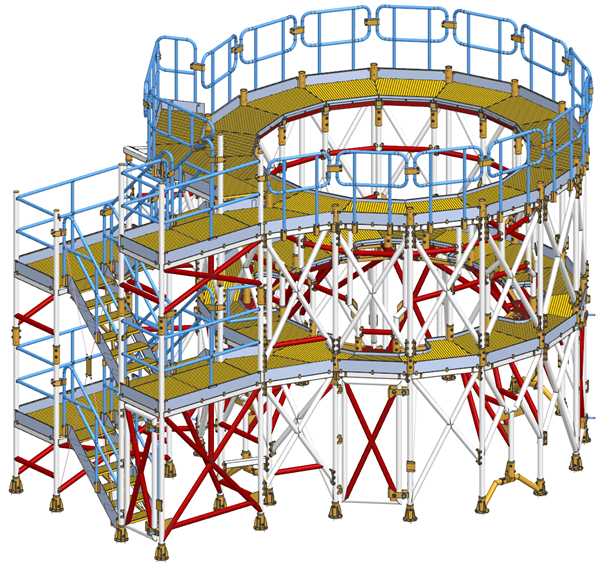}
\includegraphics[width=0.45\textwidth]{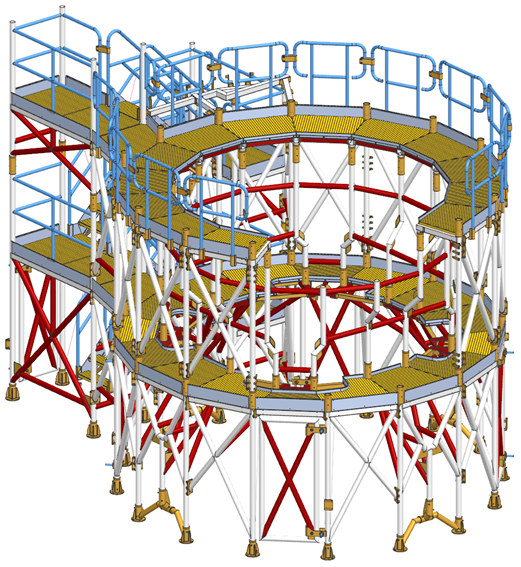}
\caption{Access Structure as Stand-Alone tool (left) and compatible with MORFEO (right).}
\label{fig:AST1}
\end{figure}

Upon movement from Nasmyth-A to Nasmyth-B platform, where MORFEO will be already installed, the extension is to be removed, and the resulting gap in the guardrail ring filled. Additionally, part of the previously fully surrounding platform is to be cut back to accommodate MORFEO’s protruding thermal duct. The newly obtained configuration will not remain unaffected over time, as several expected maintenance activities required by MORFEO demand further accommodation for components being craned in and out. These activities necessitate not only the removal of certain access structure sections but also the mounting of additional custom-made access devices such as scaffolding and step ladders (Fig.\,\ref{fig:AST1} right).

\subsubsection{Nasmyth platform constraints}
Space constraints on the ground slightly vary between Nasmyth platform A and B due to restricted zones and manufacturing tolerances. Consequently, the access structure interfaces with the ground via interchangeable reverse-engineered adapter plates of different shapes to compensate for deviations between nominal and real conditions on the platform.

Additional modularity requirements are induced by the RPA subsystem subject to access control (see Sec.\,\ref{sec:rpa}). Lockable portal segments are inserted into the access structure’s structurals for this purpose. Minor attachments such as cable trays, electric lines and plugs, emergency stop buttons, and displays necessitate additional connection features on the structure’s components.

\subsection{Finalized Access Structure Tool design}
Based on the previously described configurations, the access structure is divided into three main partitions: the staircase, which remains unchanged for every configuration, and the platform partition, where access to the instrument is provided on two elevated levels and ground level. Therefore, the platform partition divides into 1st-floor and 2nd-floor structures.

On the platform’s ground view, it is divided into 16 segments, following the access structure’s regular geometry based on 16 main columns evenly distributed over the 360° circle. The segments are numbered clockwise, beginning where the staircase is linked to the platform.

Consequently, the modular system  defines 2x16\,=\,32 distinctly denominated platform segments, characterized by their permanent use or temporary removal. These segments are grouped together into sections removed as a whole, defined by the respective maintenance situation and its geometric details.

\begin{figure}[h]
\centering
\includegraphics[width=0.44\textwidth]{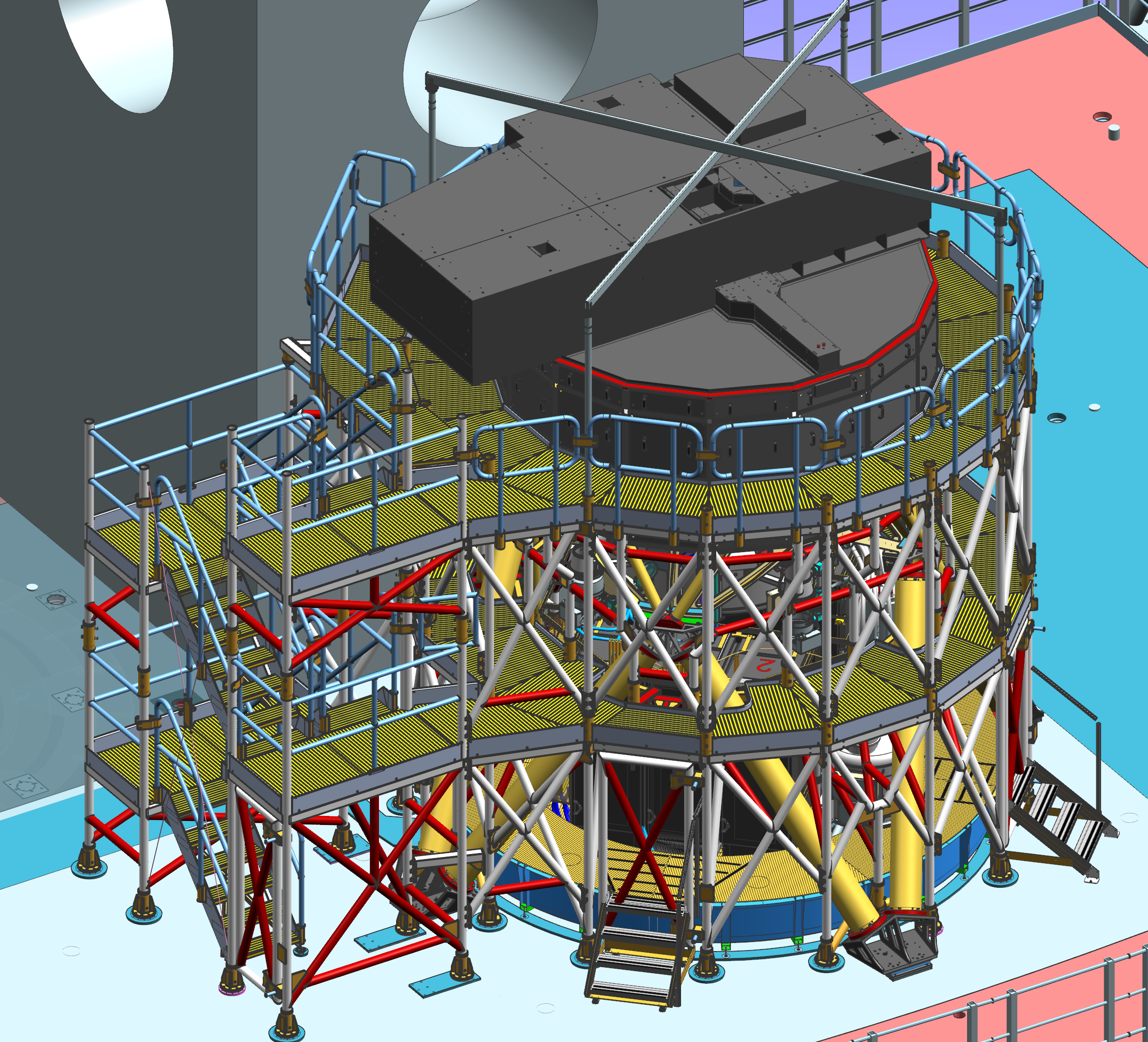}
\includegraphics[width=0.53\textwidth]{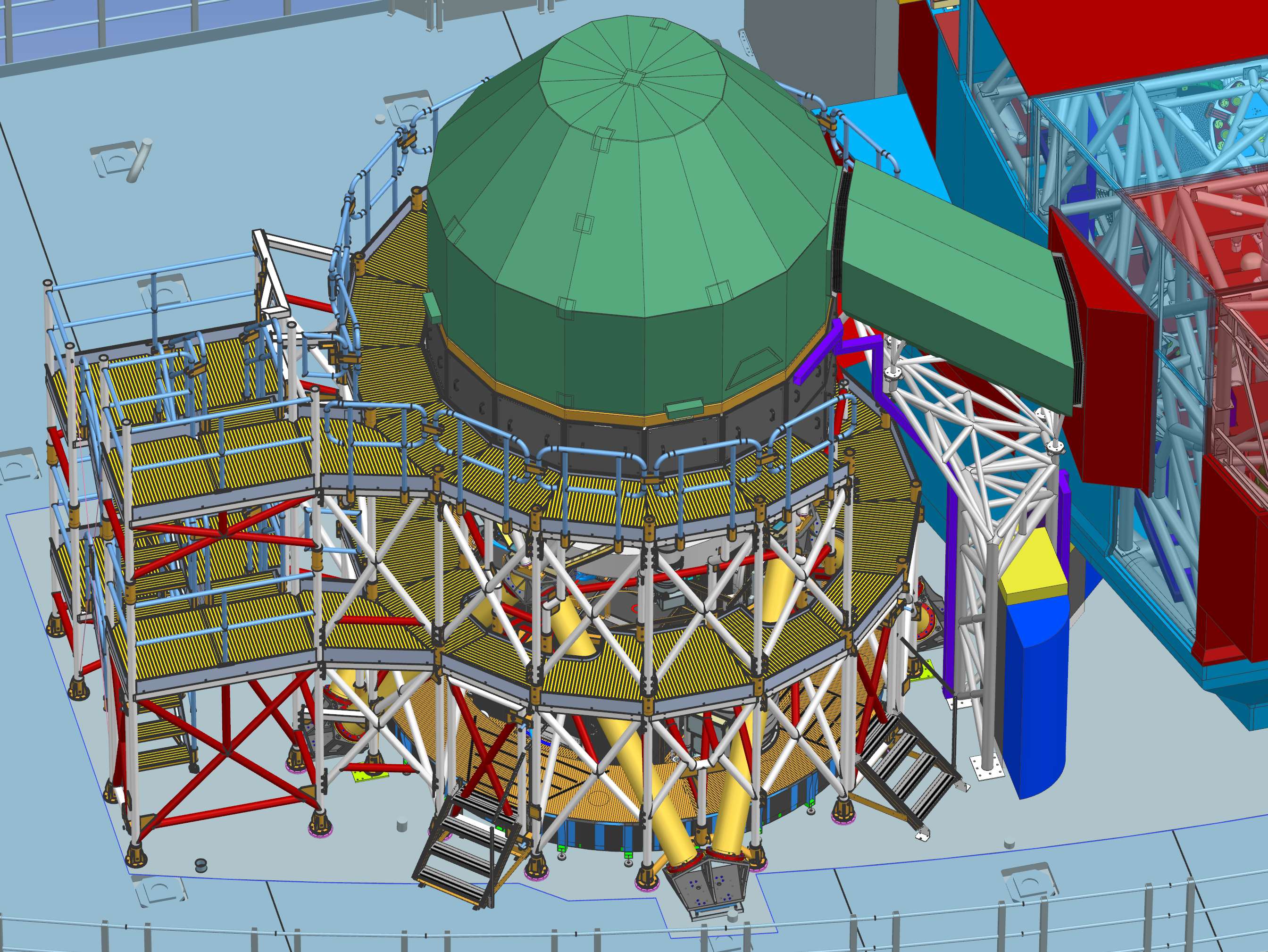}
\caption{MICADO with its Access Structure and Relay Optics on top in Stand-Alone mode (left) and in operation mode with MORFEO (right)}
\label{fig:MIC1}
\end{figure}

\subsubsection{The modular approach}
The AST's layout is based on 16 vertical main columns. Each column consists of two pipes, one for each floor, stacked via plug-in sockets. The columns have flanges welded to them where horizontal and diagonal beams engage, forming the overall trusswork and adding stiffness and stability.

Outer horizontal beams have additional attachment mounts welded for minor attachments. As the structure follows a regular and symmetric shape, these beams and attachments can easily be interchanged when switching configurations. Additional plug-in sockets on the outside allow for flexible mounting, dismounting, and interchanging of guardrails, while lugs provide the same possibilities to toe plates.

The vertical columns are plugged into feet flanged to the Nasmyth Platform via interchangeable adapters, allowing for easy switching between Stand-Alone and MORFEO modes.

\section{Vibration loads on the ELT telescope}
\label{sec:vib}
The performance of the new class of extremely large telescopes is compromised by vibrations transferred through their optics and opto-mechanics, introduced by both telescopic and instrumental subsystems. While thermal loads have traditionally been the primary concern in the design of instruments for large telescopes such as the VLT, vibrations introduce an additional layer of complexity for the ELT. This issue has gained attention as it affects the VLT and the Very Large Telescope Interferometer, particularly in their interferometric modes. A qualitative assessment is therefore essential before any quantitative evaluation.

\subsection{Qualitative approach}
Among the three subsystems discussed in this document, two are passive: the main instrument support structure (Sec.\,\ref{sec:st0}) and the access structure (Sec.\,\ref{sec:ast}). However, vibrations from the co-moving rotary platform need careful consideration:

The internal wheels of the cryogenic vessel feature notch mechanisms with minimal mass, making their impact negligible. Similarly, the support in the de-rotating system did change from a hydrostatic-oil-support into a more simple, no-pumps smooth running and robust ball-bearing. Due to the pre-load built into the ball-bearing and similar to the central bearing of the co-rotating platform, a very smooth movement and thus no vibration is to be expected from those two items. 

The most significant potential source of vibration within the co-moving rotary platform is the movement of the cable wrap (CW) on its base. Its dynamical behavior, particularly the nonlinear stick-slip effect of the four energy chains which is of non-linear characteristic,  presents the greatest uncertainty. During telescope tracking, the rotation speed is very slow, dictated by the telescope's azimuth speed and thus on its zenith distance (ZD). We assumed two rotation speeds as typical for telescope and the de-/co-rotator: The maximum tracking velocity at the edge of the zenith's blind-spot for alt-az telescopes at ZD3\textdegree \,and the general condition of ZD\,30\textdegree \,at airmass\,2. Both speeds deliver typical time constants due to the linear dimension of a single segment of each chain when passing over the gap between the CW's stationary part and its rotary housing thus leading to a passage every 35\,seconds (i.e. 0.03\,Hz) at ZD3\textdegree \,and every 300\,sec (i.e. 0.003\,Hz) at typical ZD30\textdegree \,tracking speed. Both characteristic frequencies fall outside of the three frequency windows that are specified as applicable (see Section\,\ref{sec:vib2}). 

Other considerations about the ramping-up time along the chamfer size of the stationary part lead to frequencies that fall far below the 1\,Hz lower limit of the applicable frequency domains. However, all these frequency considerations cannot fully account for the excitation that occurs when a single segment's foot strikes the chamfer of the opposing housing after crossing the gap. Therefore, a more thorough examination of this is warranted and will be addressed in the following section.

\subsection{Quantitative approach - deeper insight through a toolkit process} 
\label{sec:vib2}
The assessment discussed above, concerning its plausibility and coverage, underwent intensive scrutiny. A collaborative approach with all subsystems involved (including the Nasmyth platform itself) is recommended as the most suitable method to gain deeper quantitative insights into the emission of forces within specific frequency windows, contrasting the prior qualitative analysis. Over many years, ESO has developed a comprehensive chain of analyses, documented in numerous publications, with \cite{Sedghi2016} and \cite{Sedghi2022} serving as essential references on this topic.

From the preceding discussion, the co-moving rotary platform and its associated payload components remain as the primary potential sources of vibration. Identified candidates for vibrational excitation include:

\begin{itemize}
    \item Six electronic cabinets, including their cooling fans and air circulation systems.
    \item Two emergency and maintenance vacuum pumps.
    \item Cooling water distribution boxes.
    \item Electrical power distribution boxes.
    \item LN2 phase separator, including N2 gas recovery systems.
    \item Rotational motion of the heavy platform.
    \item Earthquake retaining system, replacing the previous roller bogies on a raceway.
    \item Drive chain, including control and servo loops.
    \item Cable wrap housing and their mounts, along with simplified energy chains.
    \item Central bearing, specifically if it has insufficient pre-load.
\end{itemize}

The pumps, valves, and other components of the distribution boxes and the LN2 phase separator are analyzed separately from the main structure. They are typically isolated from the platform through specialized damping fixtures. However, our focus will be primarily on the rotating platform and the co-moving cable wrap, as these are less amenable to isolation. This comprehensive analysis will be a collaborative effort involving IAG, MPE, and ESO. The Göttingen Astrophysics team will initiate this process by providing CAD designs and detailed specifications for the expected drive and control systems. 

Subsequently, MPE will develop a simplified finite element analysis model that includes the dynamic properties of the system, particularly focusing on the drive and servo control loops. This model will be transferred to ESO experts, who will integrate this data into their established analytical framework, refined over many years of research and application.

The next step, following the finalization of the MICADO installation design, involves constructing a detailed finite-element model from the CAD designs. This model will consider not only the mass loads but also specific drive system characteristics, such as gear stiffness and transmission properties. Importantly, the dynamic analysis will require the calculation of a large number of eigenmodes\,—\,approximately 15,000, accounting for about 30,000 states in the case of the ELT\,—\,to adequately cover the necessary frequency domain. Each of the roughly 800 mirror segments, along with their piston-tip-tilts and interface motions of other units within the telescope, must be analyzed for six degrees of freedom. These components are then correlated with the sensitivity matrix of the optical model and integrated with the primary control loops, including main axes servos, field stabilization, and adaptive optics loops. This multidisciplinary strategy culminates in a sensitivity response characterized by spatial modes and frequencies, forming a comprehensive analytical chain as outlined below:

$$ FEA + Optics + Control\,\,loops \rightarrow Forces \rightleftharpoons Wavefront\,\,error \rightleftharpoons Budgeting $$

\begin{figure}[h]
    \centering
    \includegraphics[width = 0.7 \textwidth]{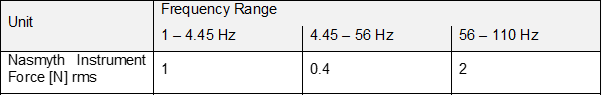}
    \caption{Allocated vibration force allowance for instruments on the Nasmyth platforms from top level requirements of the ELT.}
    \label{fig:Vibratspec}
\end{figure}

The primary mirror, comprising approximately 800 segments, is identified as the most vibration-sensitive subsystem. The allocation of vibration forces across various sub-unents of the telescope is constrained by the maximum permissible wavefront error of 50 nm. This budget allows the instruments on the Nasmyth platforms to experience vibration forces ranging from 0.4 to 2\,Newtons, depending on the frequency window (refer to Figure\,\ref{fig:Vibratspec}). 

Initial design enhancements, such as the incorporation of a retaining system at the peripheral support of the platform, have already been implemented and thus unload the central bearing. Despite these improvements, certain unknowns and uncertainties remain, which will need to be addressed in subsequent analyses. Concurrently, a comprehensive plan for testing, measuring, and verifying the platform's performance on the actual hardware is being developed.

\clearpage
\section{Earthquake loads on the MICADO camera and the ELT telescope}
\label{sec:eq}
%\subsection{Statics and Dynamics of Subsystems}
All subsystems of MICADO were analyzed for their static and dynamic performance. The dynamic performance, in particular, determines the loads imposed on the telescope's structural beams. Therefore, only the three subsystems that are directly attached to the Nasmyth platform are outlined more thoroughly here.

\subsection{The load variation according to dynamical properties}
\begin{figure}[h]
    \centering
    \includegraphics[width = 0.45 \textwidth]{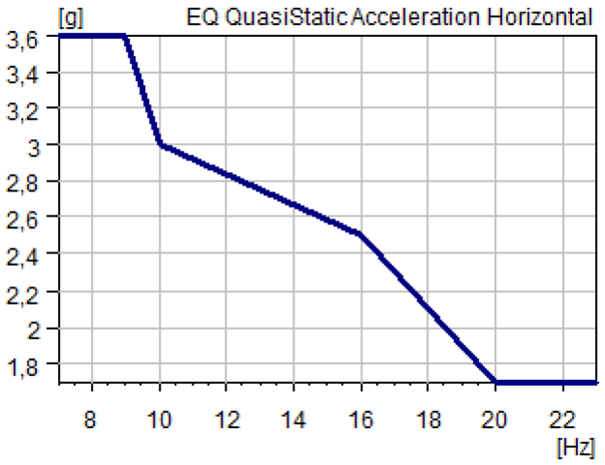}
    \includegraphics[width = 0.45 \textwidth]{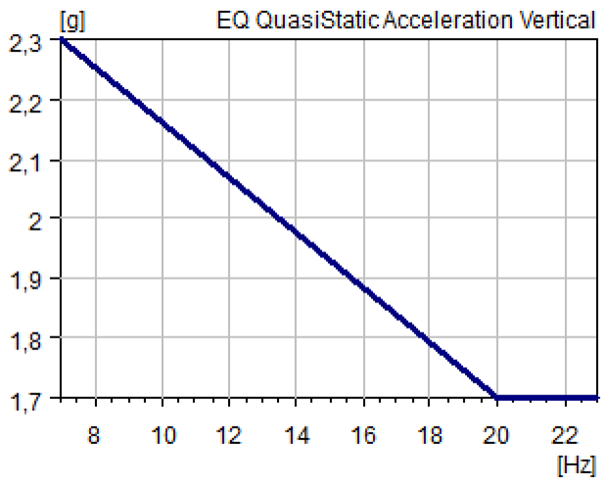}
    \caption{Applicable quasi-static horizontal (left) and vertical (right) accelerations in [g]}
    \label{fig:FEA_ESO_Fre}
\end{figure}

In the early stages of the project, the design and dynamic analysis of the telescope were so preliminary and coarse that an acceleration amplification up to 3.6\,g was specified for instruments mounted on the ELT's Nasmyth platforms. Since then, the ELT's design and modeling have become much more mature and detailed, enabling the specification of accelerations based on the lowest first eigenfrequency of the instrumental system. The load vectors for the quasi-static analysis are divided into horizontal and vertical components (see Fig.\,\ref{fig:FEA_ESO_Fre}). The horizontal loads are significantly larger than the vertical ones, reflecting the dynamical response of the telescope, with its large Nasmyth platforms, to a ground-shaking earthquake. The figure also clearly shows the reduction from the original 3.6\,g global acceleration vector to only half this value or less when the lowest eigenfrequency reaches 20\,Hz or beyond for the total instrumental system.

\subsection{The dynamics of three subsystems}
%\subsubsection{Main Support Structure}
The \textbf{main instrument structure }(see Section\,\ref{sec:st0}) supports the major components of the MICADO camera, including the cryogenic vessel with its mechanical and optical components, as well as the single and multi-conjugate AO wavefront sensors, all integrated into the mechanical de-rotator subsystem. The FEA modeling, accounting for all these subsystems, totals approximately 13\,tonnes. This comprehensive consideration led to the notable finding of a lowest eigenfrequency of about 14\,Hz for the camera system. Consequently, the net load vector, combining horizontal and vertical loads, amounts to 3.2\,g, as is summarized in the following list:

\begin{itemize}
    \item FEA modeling approach in Ansys utilizing both 3D volume and 2D shell elements.
    \item Modeled system mass - 13.2\,tonnes 
    \item Lowest estimated eigenfrequency - 13.7\,Hz
    \item Total applicable load vector - 3.2\,g acceleration.
\end{itemize}

\begin{figure}[h]
    \centering
    \includegraphics[width = 0.6 \textwidth]{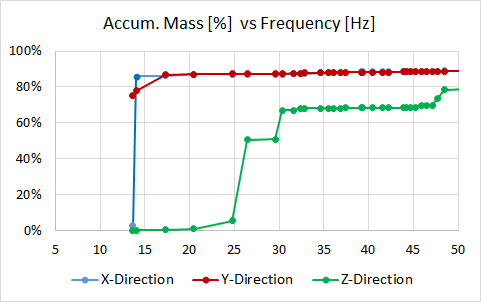}
    \caption{Eigenfrequency versus accumulated mass for the main support structure.}
    \label{fig:FEA_ST0_Fre}
\end{figure}

The dynamical behavior of the central camera system and its support structure is illustrated by the translational mode shapes depicted in Figure\,\ref{fig:FEA_ST0_Fre}. These mode shapes represent the first 30 modes, categorized according to their frequency. This visualization helps in understanding how the camera system reacts to different frequencies of vibration.

%\subsubsection{Rotary Platform}
The \textbf{rotary platform}, which synchronously co-rotates with the mechanical de-rotator of the camera (see Section\,\ref{sec:rpa}), is modeled in Ansys using a combination of volume, shell, and contact elements. These elements represent the stiff underlying structural framework and the rotary motion system. The total mass of the platform is approximately 4 tonnes, which results in an eigenfrequency close to the 7 Hz threshold. This threshold allows for a quasi-static approach in estimating earthquake loads due to mechanical decoupling between the instrument and the Nasmyth platform. The relatively low frequency is primarily driven by the six electronic cabinets that control the camera. These cabinets must move synchronously with the rotation of the camera components in its cryogenic vessel. Enhancements in earthquake resistance seem achievable for each individual cabinet. The main properties of the entire assembly are listed below:

\begin{itemize}
    \item FEA modeling approach in Ansys utilizing both 3D volume and 2D shell elements.
    \item Modeled system mass - 3.8\,tonnes 
    \item Lowest estimated eigenfrequency - 6.4\,Hz 
    \item Total applicable load vector - 4.1\,g acceleration.
\end{itemize} 

\begin{figure}[!h]
    \centering
    \includegraphics[width = 0.55 \textwidth]{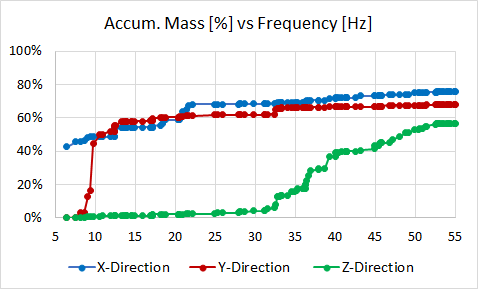}
    \caption{Eigenfrequency versus accumulated mass for the co-moving rotary platform.}
    \label{fig:FEA_RPA_Fre}
\end{figure}

The dynamical behavior of the co-rotating platform assembly, including its payload, is depicted through the translational mode shapes. These shapes are plotted in Figure\,\ref{fig:FEA_RPA_Fre} for the first 140 modes, categorized according to their frequency.

\begin{figure}[h]
    \centering
    \includegraphics[width = 0.55 \textwidth]{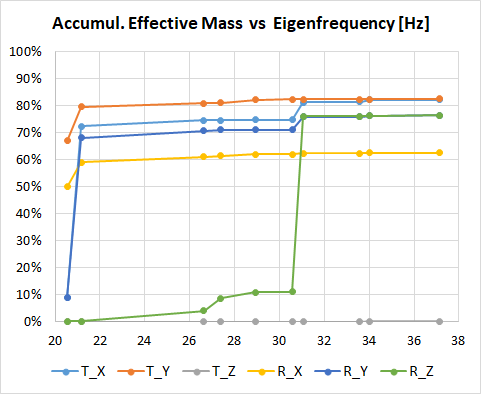}
    \vspace{0.2cm}
    \caption{Eigenfrequency versus accumulated mass for the access structure tool. T\_X, T\_Y and T\_Z refer to the translation and R\_X, R\_Y and R\_Z to rotation.}
    \label{fig:FEA_AST_Fre}
\end{figure}

%\subsubsection{Access Structure Tool}
The \textbf{access structure} (see Section\,\ref{sec:ast}), essential for accessing the entire camera installation, is the stiffest subsystem despite being designed under stringent constraints related to mass, volume, and requirements for servicing and maintenance. Nevertheless, it was possible to achieve an eigenfrequency plateau above 20\,Hz, resulting in a minimized load vector of 2.8\,g for the combined horizontal and vertical accelerations. The main properties of the assembly are listed below:
\vspace{0.5cm}
\begin{itemize}
    \item FEA modeling in Femap - in total 675\,k 2D combined with 25\,k 3D elements. 
    \item Modeled system mass - 3.0\,tonnes 
    \item Lowest estimated Eigenfrequency - 20.5\,Hz 
    \item Total applicable load vector - 2.8\,g acceleration. 
\end{itemize} 
\vspace{1cm}
The dynamical behavior of the access structure tool, used for maintenance and repair, is depicted through both translational and rotational mode shapes. These shapes are plotted in Figure\,\ref{fig:FEA_AST_Fre} for the first 10 modes, categorized according to their frequency in the range of 21 to 40 Hz. 
\vspace{1cm}

\clearpage
\subsection{Nasmyth platform structure loads}

\begin{figure}[h]
    \centering
    \begin{subfigure}[b]{0.9\textwidth}
        \centering
        \includegraphics[width=\textwidth]{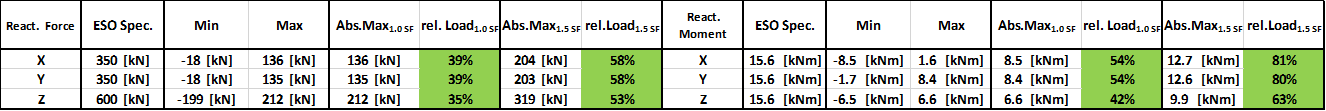}
        \caption{Support Structure induced forces/moments.}
        \label{subfig:FEA_ST0_RFM}
    \end{subfigure}
    \begin{subfigure}[b]{0.9\textwidth}
        \centering
        \includegraphics[width=\textwidth]{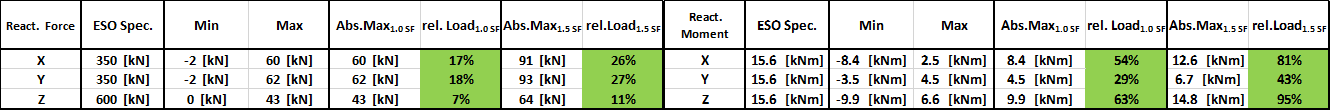}
        \caption{Rotary Platform induced forces/moments.}
        \label{subfig:FEA_RPA_RFM}
    \end{subfigure}
    \begin{subfigure}[b]{0.9\textwidth}
        \centering
        \includegraphics[width=\textwidth]{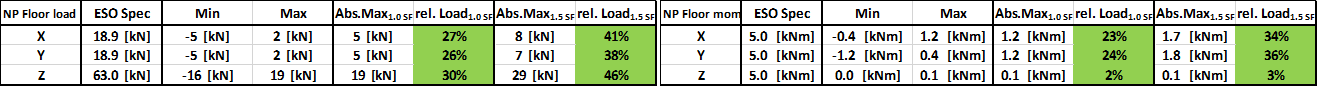}
        \caption{Access Structure induced forces/moments.}
        \label{subfig:FEA_AST_RFM}
    \end{subfigure}
    \caption{Reaction forces/moments onto Nasmyth platform by the three different subsystems in absolute value and in percentage of allowed capacities.}
    \label{fig:RFM_combined}
\end{figure}

The dynamics of the three subsystems outlined in the previous subsection determines the applicable loads for each and thus the resulting reaction forces and moments that act onto their respective attachment. As two different types of attachment are used on the Nasmyth platform -~specific high-capacity attachment points and the Nasmyth floor itself~- two different capacity limits has to be met; 350\,kN (600\,kN\,vertical) and 15\,kNm in the high-capacity case and 19\,kN (63\,kN\,vertical) and 5\,kNm for the Nasmyth floor in the coordinate system's three main axes (cf.\,Figure\,\ref{fig:RFM_combined} first column). The reaction forces and moments are estimated with the FEA models described above. On top of these estimates a safety factor of SF\,1.5 had to be applied. Those and the net reaction loads in force and moments are plotted for all the three subsystems in Figure\,\ref{fig:RFM_combined} including the used capacity percentage thus demonstrating the remaining contingency.  

% \begin{figure}
%     \centering
%     \includegraphics[width = 0.9 \textwidth]{Pictures/FEA_ST0_ReactFM.png}
%     \caption{Reaction forces/moments onto Nasmyth platform through the main Support Structure}
%     \label{fig:FEA_ST0_RFM}
% \end{figure}

% \begin{figure}
%     \centering
%     \includegraphics[width = 0.9 \textwidth]{Pictures/FEA_RPA_ReactFM.png}
%     \caption{Reaction forces/moments onto Nasmyth platform through the Rotary Platform Assembly}
%     \label{fig:FEA_RPA_RFM}
% \end{figure}

% \begin{figure}
%     \centering
%     \includegraphics[width = 0.9 \textwidth]{Pictures/FEA_AST_ReactFM.png}
%     \caption{Reaction forces/moments onto Nasmyth platform floor through the Access Structure Tool}
%     \label{fig:FEA_AST_RFM}
% \end{figure}

%\section{Summary or Conclusion ??}

%\input{summary}

\section*{Ackowledgments}

This project is supported by the German Federal Ministry of Education and Research BMBF under the research grants 05A11MG1, 05A14MG2, 05A17MG1, 05A20MG2 and 05A23MG2. We also thank Babak Sedghi and his collaborators at ESO for providing us the insights into their analysis chain on vibrational assessment that they developed over many years.

\bibliography{SPIE2024} % bibliography data in report.bib

@inproceedings{Sedghi2022,
author = {B. Sedghi and P. Zuluaga Ramirez and D. Pilbauer and S. Leveratto and U. Lampater and M. M{\"u}ller and G. Jakob and M. Haug and M. Accardo and Y. Lammen and J. Abad-Pastor and J. C. Gonz{\'a}lez-Herrera},
title = {{ESO ELT: vibration performance and budget verification: measured equipment data as input to telescope model}},
volume = {12187},
booktitle = {Modeling, Systems Engineering, and Project Management for Astronomy X},
editor = {George Z. Angeli and Philippe Dierickx},
organization = {International Society for Optics and Photonics},
publisher = {SPIE},
pages = {1218714},
year = {2022},
doi = {10.1117/12.2628907},
URL = {https://doi.org/10.1117/12.2628907}
}

@inproceedings{Sedghi2016,
author = {B. Sedghi and M. M{\"u}ller and G. Jakob},
title = {{E-ELT vibration modeling, simulation, and budgeting}},
volume = {10012},
booktitle = {Integrated Modeling of Complex Optomechanical Systems II},
editor = {Marco Riva},
organization = {International Society for Optics and Photonics},
publisher = {SPIE},
pages = {1001202},
year = {2016},
doi = {10.1117/12.2200926},
URL = {https://doi.org/10.1117/12.2200926}
}
\bibliographystyle{plainnat} % makes bibtex use spiebib.bst
\end{document}